\documentclass{epl}

\def\d{{\rm d}}

\title{Capillary-gravity waves: a ``fixed-depth" analysis.}
\author{F. Chevy \and E. Rapha\"{e}l}
\institute{ Laboratoire de Physique de la Mati\`ere Condens\'ee,
UMR CNRS 7125 and FR CNRS 2438, Coll\`ege de France 75231 Paris
Cedex 05, France }


\begin{document}

\maketitle

\begin{abstract}
We study the onset of the wave-resistance due to the generation of
capillary-gravity waves by a partially immersed moving object in
the case where the object is hold at a fixed immersion depth. We
show that, in this case, the wave resistance varies continuously
with the velocity, in qualitative accordance with recent
experiments by Burghelea {\em et al.} (Phys. Rev. Lett. {\bf 86},
2557 (2001)).
\end{abstract}

\section{Introduction}

The dispersive properties of capillary-gravity waves are
responsible for the complicated wave pattern generated at the free
surface of a still liquid by a disturbance moving with a velocity
$V$ greater than the minimum phase speed $V_{\rm
c}=\left(4g\gamma/\rho\right)^{1/4}$, where $g$ is the gravity,
$\gamma$ is the surface tension  and $\rho$ the density of the
fluid \cite{Lighthill78}.  The disturbance may be produced by a
small object partially immersed in the liquid or by the
application of an external surface pressure distribution
\cite{Kelvin}. The waves generated by the moving perturbation
propagate momentum to infinity and, consequently, the disturbance
 experiences
 a drag $\bf R$ called the wave resistance
\cite{Debnath94}. For $V<V_{\rm c}$ the wave resistance is equal
to zero since, in this case, no propagating long-range waves are
generated by the disturbance \cite{Stokes}.

A few years ago, it was predicted that the wave resistance
corresponding to a surface pressure distribution symmetrical about
a point should be discontinuous at $V=V_{\rm c}$ \cite{Raphael97}.
More precisely, if $F_0$ is the the total vertical force exerted
on the fluid surface, the wave resistance is expected to reach a
finite value $R_{\rm c}>0$ for $V\rightarrow V_{\rm c}^+$. For  an
object much smaller than the capillary length
$\kappa^{-1}=\sqrt{\gamma/\rho g}$, the discontinuity $R_{\rm c}$
is given by:

\begin{equation}
R_c=\frac{F_0^2}{2\sqrt{2}} \frac{\kappa}{\gamma}.
\label{PGG}\end{equation}

Experimentally,   the onset of the wave-resistance due to the
generation of capillary-gravity waves by a partially immersed
moving object was studied recently by two independent groups
\cite{Browaeys01,Steinberg}. While Browaeys {\em et al.}
\cite{Browaeys01} indeed find a discontinuous behaviour of the
wave-resistance at $V=V_{\rm c}$ in agreement with the theoretical
predictions, Burghelea {\em et al.} \cite{Steinberg} observe a
smooth transition.

The discrepancy between the theoretical analysis of
\cite{Raphael97} and the experimental results of \cite{Steinberg}
might be due to the fact that the experimental setup of Burghelea
{\em et al.} uses a feedback loop to keep the object at a constant
depth while the analysis of \cite{Raphael97} assumes that the
vertical  component  $F_0$ of the force exerted by the disturbance
on the fluid  does not depend on the velocity $V$ (we might call
such an analysis a ``fixed force" analysis). In order to check
this proposition, we perform in this letter a ``fixed-depth"
calculation of the wave-drag close to the onset threshold. A
somewhat  similar analysis was performed in the large velocity
limit by Sun and Keller \cite{Sun01}. We will show that such a
calculation indeed yields a cancellation of the vertical force at
$V=V_{\rm c}$, that is to say, according to equation (\ref{PGG}),
a smoothing of the discontinuity.

\begin{figure}
\centerline{\includegraphics{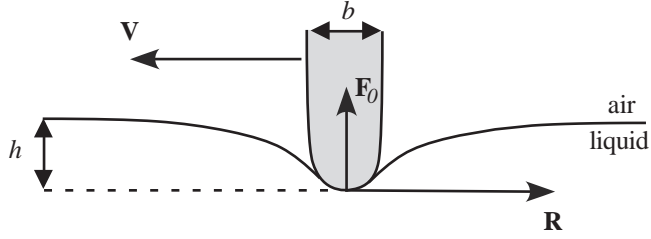}}
 \caption{We study the
behavior of a hydrophobic  moving object of characteristic size
$b$ immersed at a depth $h$. The velocity of the object is $V$. We
decompose the force exerted by the fluid on the object in a
component orthogonal to the free surface ($F_0$) and a component
parallel to it ($R$).} \label{Fig0}
\end{figure}

\section{Model}

We take the $x,y$ plane as the equilibrium surface of the fluid.
The immersed object exerts a stress at the fluid surface that can
be considered  equivalent to a pressure field $p$
\cite{SurfaceForce} that travels over the surface with a velocity
$V$ in the $x$ direction. We assume $\hat p$ (the Fourier
transform of $p$) to be of the form:

\begin{equation}
\hat p (k_x,k_y)=F_0 \hat\phi (k), \label{Ansatz}
\end{equation}

\noindent  where $k=\sqrt{k_x^2+k_y^2}$ and $\hat\phi (0)=1$. In
this case, $\hat p$ is isotropic \cite{Isotropic} and $F_0$ is the
total vertical force exerted on the fluid.

Within the framework of  Rayleigh's linearized theory of
capillary-gravity waves,  the Fourier transform $\hat\xi ({\mathbf
k})$ of the free surface displacement $\xi ({\bf r})$ is related
to the pressure field through \cite{Richard99}:

\begin{equation}
\hat\xi (k_x,k_y)=-F_0\frac{k}{\rho}\left(\frac{\hat\phi
(k)}{\omega_0^2 (k)-4\nu^2k^3 q+(2\nu k^2-i{\mathbf
V}\cdot{\mathbf k})^2}\right), \label{LinearReponse}
\end{equation}

\noindent where $\omega_0^2 (k)=gk+\gamma k^3/\rho$ is the free
dispersion relation, $q^2=k^2-i{\mathbf k}\cdot{\mathbf V}/\nu$
and $\nu=\eta/\rho$ is the kinematic viscosity of the fluid.

Let us suppose the object is located at the origin of the moving
frame. If $h$ is its depth, the free surface displacement $\xi$
must be $-h$ under the pinpoint (here we suppose the object
sufficiently  hydrophobic and $h$ not to large so that the
pinpoint does not pierce  the surface). This leads to the
following normalization condition:

\begin{equation}
\xi ({\mathbf 0})=\int \frac{\d^2k}{(2\pi)^2}\,\hat\xi ({\mathbf
k})=-h. \label{Boundary}
\end{equation}

The value of $h$ as a function of $F_0$ can be readily calculated
using equations (\ref{LinearReponse}) and (\ref{Boundary}):

\begin{equation}
h=\Xi (V) F_0, \label{FormalDepth}
\end{equation}

\noindent with

\begin{equation}
\Xi
(V)=\int\frac{\d^2k}{(2\pi)^2}\frac{k}{\rho}\left(\frac{\hat\phi
(k) }{\omega_0^2(k)-4\nu^2k^3 q^2+(2\nu k^2-i{\mathbf
V}\cdot{\mathbf k})^2}\right). \label{Normalize}
\end{equation}

Finally, the drag-force $\mathbf R$  is calculated by simply
integrating the pressure force over the free surface
\cite{Havelock10}.

\begin{equation}
{\mathbf R}=-\int \d^2r\, p({\mathbf r}){\mathbf\nabla}\xi
({\mathbf r})=-\int \frac{\d^2k}{(2\pi)^2}\, i{\mathbf k}\hat
p^*({\mathbf k})\hat\xi ({\mathbf k}).
\end{equation}

This yields, using the explicit expression (\ref{LinearReponse})
for $\hat\xi$,

\begin{equation}
{\mathbf R}=F_0^2{\mathbf \Lambda} (V), \label{FormalDrag}
\end{equation}

\noindent with

\begin{equation}
{\mathbf \Lambda} (V)=\int
\frac{\d^2k}{(2\pi)^2\rho}\left(\frac{ik{\mathbf k}|\hat\phi
(k)|^2}{\omega_0^2(k)-4\nu^2k^3 q^2+(2\nu k^2-i{\mathbf
V}\cdot{\mathbf k})^2}\right). \label{DragForce}
\end{equation}

According to equation (\ref{FormalDrag}), the integral $\mathbf
\Lambda$ describes the fixed-force behaviour of the
wave-resistance. Due to the symmetry of $\hat\phi$, $\mathbf
\Lambda$ is parallel to $\mathbf V$ and we shall henceforth set
${\mathbf\Lambda}=\Lambda {\mathbf u}$, where ${\mathbf
u}={\mathbf V}/V$ is the unit vector parallel to the velocity of
the object. The authors of \cite{Raphael97} studied the properties
of $\Lambda$ in the case of a non-viscous fluid for which
 they showed that:

 (a) $\Lambda=0$ for $V<V_{\rm c}$;

 (b) $\Lambda$
is discontinuous for $V\rightarrow V_{\rm c}^+$, with

\begin{equation}\lim_{V\rightarrow V_{\rm c}^+}{
\Lambda}=\Lambda_{\rm c}=\frac{1}{2\sqrt{2}}\frac{\kappa}{\gamma};
\label{LambdaThreshold}
\end{equation}

(c) in the large velocity limit

\begin{equation}
\Lambda\sim  \frac{2\rho V^2}{3\pi \gamma^2}. \label{LambdaLarge}
\end{equation}

In the case of a fixed depth analysis,  $F_0$ becomes a function
of $V$. Using equations (\ref{FormalDepth}), we can rewrite the
wave-resistance as

\begin{equation}
{\mathbf R}=h^2\frac{{\mathbf \Lambda} (V)}{\Xi^2 (V)}.
\label{FormalFixedDepth}
\end{equation}

\begin{figure}
\centerline{\includegraphics{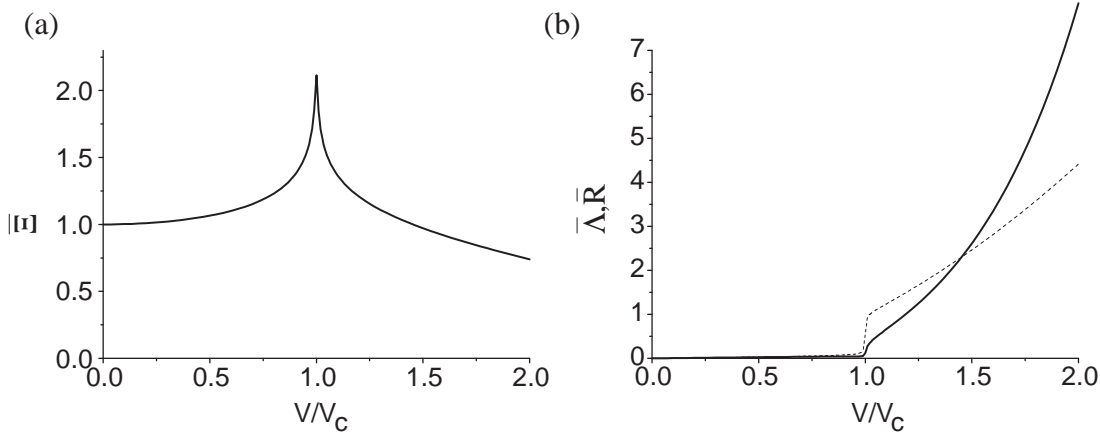}}
 \caption{(a) Numerical calculation of the  dimensionless integral $\bar\Xi=\Xi (V)/\Xi (0)$. (b) Comparison of $\bar\Lambda (V)=\Lambda (V)/\Lambda_{\rm c}$
(dashed line) and $\bar R=\bar\Lambda/\bar\Xi^2$ (full line)
describing respectively the fixed-force and fixed-depth behaviour
of the wave-resistance. We clearly see that contrary to the
fixed-force analysis, the fixed-depth calculation does not yield
any discontinuity at the threshold. Here $\nu=10^{-6}~{\rm
m^2\cdot s^{-1}}$, $\gamma=72~{\rm mN\cdot m^{-1}}$,
$\rho=1000~{\rm kg\cdot m^{-3}}$ and $b=0.1~{\rm mm}$.}
\label{fig2}
\end{figure}

In general  we have to rely on numerics to calculate the integrals
$\Xi$ and $\Lambda$. Typical results are presented on Figure
\ref{fig2} for an objet of size 0.1~mm immersed in water, and for
a step-like  function $\hat\phi$ equal to 1 for $k<1/b$ and $0$
otherwise.

We first observe on Fig. (\ref{fig2}-a) that $\Xi$ increases
sharply near the threshold. This leads to two rather different
behaviours for $\Lambda$ and $R$ as shown on Fig. (\ref{fig2}-b):
while $\Lambda$ exhibits a discontinuity close to $V=V_{\rm c}$
\cite{Viscosity}, the fixed-depth wave-drag $R=\Lambda/\Xi^2$
cancels smoothly at the critical velocity.

\section{Inviscid flow}

The characteristic features displayed by the plots of Fig.
\ref{fig2} can be captured by a zero-viscosity analysis. Setting
$\nu=0$, equation (\ref{Normalize}) can be simplified as:

\begin{equation}
\Xi={\cal P}\int\frac{\d^2 k}{(2\pi)^2}
\frac{k}{\rho}\left(\frac{\hat\phi (k) }{\omega_0^2(k)-({\mathbf
k}\cdot{\mathbf V})^2}\right), \label{ViscZero}
\end{equation}

\noindent where $\cal P$ denotes the Cauchy principal value of the
integral. The integral is calculated in polar coordinates
$(k,\theta)$, where $\theta$ is the angle of $\mathbf k$ with
respect to ${\mathbf V}$. Introducing the  function $G$ defined
by:

\begin{equation}
G(k)={\cal P
}\int_0^{2\pi}\frac{\d\theta}{2\pi}\left(\frac{1}{m_k^2-2{\cal
M}^2\cos^2 (\theta)}\right),
\end{equation}

\noindent where $m_k^2=k/\kappa+\kappa/k$ and ${\cal M}=V/V_{\rm
c}$ is the ``Mach" number,  equation (\ref{ViscZero}) can then be
rewritten as:

\begin{equation}
\Xi=\frac{1}{\gamma\kappa}\int_0^\infty\frac{\d
k}{2\pi}\hat\phi(k)G(k). \label{EqnF}
\end{equation}

\begin{figure}
\centerline{\includegraphics{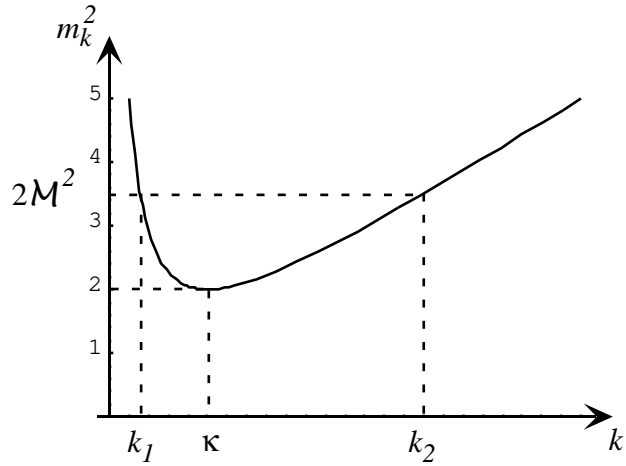}} \caption{Variations of
$m_k^2$ (see text) with k. Since $m_k^2$ reaches a minimum for
$k=\kappa$ with $m_\kappa^2=2$, equation $m_k^2=2{\cal M}^2$
presents solutions for ${\cal M}>1$. In this case, there are two
possible values $k_1$ and $k_2$ with $k_1<\kappa<k_2$.}
\label{Fig1}
\end{figure}

Using the residue theorem \cite{Riley98}, we get:

\begin{equation}
G(k)=\frac{1}{\sqrt{m_k^2(m_k^2-2{\cal
M}^2)}}\,\Theta(m_k^2-2{\cal M}^2),
\end{equation}

\noindent where $\Theta$ the is the Heaviside step-function. The
variations of  $m_k^2$ with $k$  are plotted on Fig. \ref{Fig1}:
$m_k^2$ reaches its minimum value $(m_k^2)_{\rm min}=2$ for
$k=\kappa$. It shows that equation $2{\cal M}^2-m_k^2=0$ has two
solutions $k_1$ and $k_2$, with $k_1<\kappa<k_2$, if $V$ is larger
than the critical velocity $V_{\rm c}$, and none if $V<V_{\rm c}$.

For $V>V_{\rm c}$, $\Xi$ evaluates to

\begin{equation}
\Xi=\frac{1}{\gamma\kappa}\left[\int_0^{k_1}\frac{\d
k}{2\pi}\frac{\hat\phi(k)}{\sqrt{m_k^2(m_k^2-2{\cal
M}^2)}}+\int_{k_2}^\infty\frac{dk}{2\pi}\frac{\hat\phi(k)}{\sqrt{m_k^2(m_k^2-2{\cal
M}^2)}}\right]. \label{Integral}
\end{equation}

The above integrals can be calculated in the two limiting cases
 ${\cal M}\approx 1$ ({\em i.e.} $V\approx V_{\rm c}$) and ${\cal M}\gg
1$ ({\em i.e.} $V\gg V_{\rm c}$).

For large $\cal M$,  the integrals are restricted to either large
or small values of $k$. If $\hat\phi$ vanishes faster than $1/k$
for large $k$, we can show that the small $k$ contribution
dominates. In this region, the dispersion relation is dominated by
gravity waves, so that we can approximate $m_k^2$ by $k/\kappa$. A
straightforward calculation then yields the following asymptotic
expansion for $\Xi$:

\begin{equation}
\Xi\sim\frac{1}{6\pi\gamma}\left(\frac{V_{\rm c}}{V}\right)^4.
\label{LargeM}
\end{equation}

Combining this result with (\ref{LambdaLarge}) and
\ref{FormalFixedDepth}, we get:

$$R\sim 12\pi\rho h^2\frac{V^{10}}{V_{\rm c}^8}.$$

Let's now focus on the case $V$ close to $V_{\rm c}$.  We set:

$$\widetilde\Xi_1=\int_0^{k_1}\frac{\d
k}{2\pi}\frac{\hat\phi(k)}{\sqrt{m_k^2(m_k^2-2{\cal M}^2)}},$$

and

$$\widetilde\Xi_2=\int_{k_2}^\infty\frac{\d
k}{2\pi}\frac{\hat\phi(k)}{\sqrt{m_k^2(m_k^2-2{\cal M}^2)}}.$$

Using the fact that $k_1$ and $k_2$ are the roots of $m_k^2=2{\cal
M}^2$, we can rewrite $\widetilde\Xi_1$ as:

$$\widetilde\Xi_1=\int_0^{k_1}\frac{\d
k}{2\pi} \frac{k\hat\phi(k)}{\sqrt{(k^2/\kappa^2)(k-k_1)
(k-k_2)}}.$$

 When $V\rightarrow
V_{\rm c}$, the $(k-k_1) (k-k_2)$ term cancels in $k=\kappa$.
Since all other terms are regular,  we can write at the leading
order:

$$\widetilde\Xi_1\sim\kappa\frac{\hat\phi(\kappa)}{\sqrt{2}}\int_0^{k_1}\frac{\d
k}{2\pi}\frac{1}{\sqrt{(k-k_1) (k-k_2)}}.$$

This latter integral is readily calculated and gets a very simple
form in the limit $V\rightarrow V_{\rm c}$:

$$\widetilde\Xi_1\sim\kappa\frac{\hat\phi(\kappa)}{\sqrt{2}}\ln
\left(k_2-k_1\right)\sim\kappa\frac{\hat\phi(\kappa)}{2\sqrt{2}}\ln
\left({\cal M}-1\right).$$

Since $m_k^2$ is invariant by the transformation
$k/\kappa\rightarrow \kappa/k$, we see that $\Xi_1$ and $\Xi_2$
have the same asymptotic behaviour for $V\rightarrow V_{\rm c}$.
In this limit we have
$\Xi=\left(\widetilde\Xi_1+\widetilde\Xi_2\right)/\gamma\kappa
\approx 2\widetilde\Xi_1/\gamma\kappa$, hence:

\begin{equation}
\Xi\sim \frac{1}{2\pi\sqrt{2}\gamma}\,\hat\phi (\kappa) \ln \left(
{\cal M}-1\right). \label{ApproxBoundary}
\end{equation}

In the case of an object of size $b$, the width of $\hat\phi$ is
about $1/b$. If we choose $b$ much smaller than the capillary
length $\kappa^{-1}$, we can approximate $\hat\phi (\kappa)$ by
$\hat\phi (0)=1$. In this limit, equation (\ref{ApproxBoundary})
takes the following form:

\begin{equation}
\Xi\sim \frac{1}{2\pi\sqrt{2}\gamma} \ln \left({\cal M}-1
\right)\sim \frac{1}{2\pi\sqrt{2}\gamma} \ln \left(\frac{V-V_{\rm
c}}{V_{\rm c}} \right).\label{MainResult}
\end{equation}

If we combine equations (\ref{LambdaThreshold}) and
(\ref{MainResult}), we see that slightly above the threshold, the
wave-resistance behaves like:

\begin{equation}
R\sim\frac{4\pi^2}{\sqrt{2}}\left(\frac{\gamma \kappa h^2}{\ln^2
\left( V/V_{\rm c}-1\right)}\right). \label{Scaling}
\end{equation}

 Equation (\ref{Scaling}) constitutes the main result of this paper.  First, we notice that
for small objects this relation is, as expected, independent of
the actual shape of the pressure field. Second, and more
important, it shows the wave resistance $R$ cancels out at
$V=V_{\rm c}$. This smearing is due to the cancellation of the
vertical force $F_0$ near the threshold that we get from the
behaviour of $\Xi$.

\section{Conclusion}

In this paper we have shown that in a fixed-depth situation the
discontinuity of the drag-force calculated in \cite{Raphael97}
vanishes and is replaced by a smooth variation, in accordance with
the experimental results found in \cite{Steinberg}. A quantitative
comparison between the present analysis and the data of Ref.
\cite{Steinberg} is however more involved since experiments  from
\cite{Steinberg} were performed in narrow channel geometry
\cite{Steinberg02} with objects of size comparable with the
capillary length. To recover the scaling relation observed
experimentally, it would also be necessary to take into account
both the variations of the shape of $\hat\phi$ with $V$ and the
reflections of the waves on the walls of the channel. The present
study suggests nevertheless that to fully test relation
(\ref{PGG}), experiments need to be devised that would measure
both $R$ and $F_0$.

\acknowledgments We wish to thank J. Browaeys, P.-G. de Gennes and
D. Richard for very helpful discussions, as well as V. Steinberg
for sending us his experimental data prior to publication.


\begin{thebibliography}{30}
\bibitem{Stokes} Note that for a viscous fluid, the moving perturbation
is also subjected to the usual Stokes drag, for both $V<V_{\rm c}$
and $V>V_{\rm c}$.
\bibitem{SurfaceForce} Not that this  assumption is not entirely correct in the
presence of viscosity since it leads to a slip of the object with
respect to the fluid surface. In this case, a force field parallel
to the free surface should be added to take into account the
viscous drag. However, this new term leads to the well known
Stokes drag, that can substracted from experimental results, and
so can be omitted in our analysis.
\bibitem{Isotropic} This latter
hypothesis is not too restrictive in the case of very small
objects for which the actual shape is not relevant.
\bibitem{Viscosity} For an inviscid fluid, this increase is actually a discontinuity. For high viscosities, we have observed no noticeable
accident of $\mathbf \Lambda$ for $V$ close to $V_{\rm c}$.
\end{thebibliography}
\end{document}